\documentclass{iau}

\usepackage{amsmath}
\usepackage{graphicx}
\usepackage{multirow}
\usepackage{amssymb}
\usepackage{hyperref}
\usepackage{listings}
\usepackage{xcolor}
\usepackage{natbib}

\lstdefinelanguage{SQL}{
  keywords={SELECT, FROM, WHERE, AND, OR, JOIN, ON, TOP, INTO, VALUES, CREATE, TABLE, PRIMARY, KEY, FOREIGN, NOT, NULL, INNER, OUTER, LEFT, RIGHT},
  sensitive=false,
  morecomment=[l]{--},
  morestring=[b]',
}
\begin{document}

\lefttitle{Prathamesh Tamhane}
\righttitle{NL-to-SQL for data retrieval}

\jnlPage{1}{7}
\jnlDoiYr{2025}
\doival{10.1017/xxxxx}

\aopheadtitle{Proceedings IAU Symposium}
\editors{I. Liodakis, eds.}

\title{A Natural Language Interface for Efficient Data Retrieval in SDSS}

\author{Prathamesh Tamhane}
\affiliation{Department of Physics and Astronomy, University of Alabama in Huntsville, 301 Sparkman Drive, Huntsville, AL 35899, USA \email{pdt0003@uah.edu}}

\begin{abstract}
Modern astronomical surveys such as the Sloan Digital Sky Survey (SDSS) provide extensive astronomical databases enabling researchers to access vast amount of diverse data. However, retrieving data from archives requires knowledge of query languages and familiarity with their schema, which presents a barrier for non-experts. This work investigates the use of Microsoft Phi-2, a compact yet powerful transformer-based language model, fine-tuned on natural language--SQL pairs constructed from SDSS query examples. We develop an interface that translates user queries in natural language into SQL commands compatible with SDSS SkyServer. Preliminary evaluation shows that the fine-tuned model produces syntactically valid and largely semantically correct queries across a variety of astronomy-related requests. Our results show that even small-scale models, when carefully fine-tuned, can provide effective domain-specific natural language interfaces for large scientific databases.
\end{abstract}


\maketitle

\section{Introduction}

Modern astronomical surveys provide vast archives of photometric, spectroscopic, and other types of data, often organized across complex tables and views. For example, the Sloan Digital Sky Survey \citep[SDSS;][]{york00}, which has transformed modern astronomy by providing one of the largest and most influential data archives, has data spanning hundreds of tables and views. Upcoming surveys such as LSST, DESI, Euclid, the Roman Space Telescope and SKA will further expand these multi-dimensional, large-scale datasets, increasing both their scientific potential and the complexity of accessing and querying them efficiently. Accessing these data typically requires proficiency in SQL or equivalent query languages as well as an understanding of complex database schemas, creating a barrier for many astronomers, especially students, educators, and researchers at smaller institutions who may not have the time or resources to develop these specialized skills. Although such surveys provide interfaces to explore these datasets, the underlying challenges of navigating complex schemas remain common across many surveys.

Natural Language Interfaces to Databases (NLIDBs) offer an alternative by allowing users to formulate queries in plain language, which are then automatically translated into SQL. This approach has been explored in computer science as a means to allow nonexperts to query complex databases without specialized programming knowledge \citep[e.g.,][]{zelle96,li14,yu19}. In astronomy, such interfaces remain uncommon. However, these tools offer clear utility for preliminary exploratory studies and for educational purposes, enabling students or new researchers to engage with real datasets without mastering complex query languages. Only recently advances in large language models (LLMs) have made NLIDBs practical at scale. Models such as GPT-4 have demonstrated remarkable ability to parse and reason over natural language. However, they are computationally expensive and rely on large-scale infrastructure making them difficult to deploy in domain-specific applications like astronomy, where lightweight, adaptable solutions are more practical. Since several LLMs have already been pre-trained and released as open-source resources, it is no longer necessary to train models entirely from scratch. Instead, our objective is to investigate whether fine-tuning an existing pre-trained model can achieve strong performance on domain-specific tasks in astronomy. In this work, we explore fine-tuning Microsoft Phi-2 \citep{javaheripi23}, a relatively small 2.7 billion parameter transformer model with strong reasoning capabilities, to serve as a natural language interface tailored to SDSS. Our goal is to lower the technical barrier to data access, enabling astronomers to explore SDSS data more efficiently and intuitively.

\section{Data and Methodology}

To develop a domain-specific model, we constructed a dataset of natural language (NL) queries paired with corresponding SDSS SQL statements. We began with example queries from the SDSS SkyServer and CasJobs tutorials, which are widely used in training and outreach, and manually created a small initial set. These were then expanded through natural language rewrites that preserved the same intent but used different phrasings to increase linguistic diversity. To further augment the dataset, we employed paraphrase generation via prompt-based rephrasing with larger LLMs such as ChatGPT, yielding a broader range of query formulations. In addition, we used a script-based approach to generate synthetic training examples. The script randomly samples SDSS parameters such as colors, magnitudes, redshift ranges, object types, extinction values, and concentration indices, and combines multiple conditions to create diverse queries. Each selected parameter is randomly assigned a value within a physically meaningful range. The script also selects additional output columns when needed (e.g., extinction or concentration indices) to ensure the SQL query returns all relevant data. A natural language instruction is automatically constructed from templates describing the selected conditions, while the corresponding SQL query is built with \texttt{PhotoObj} and \texttt{SpecObj} joins, applying the same filters.  

For the corresponding SQL queries, we joined the \texttt{PhotoObj} and \texttt{SpecObj} views in SDSS and always returned $u$--$i$ band magnitudes, RA/Dec coordinates, and redshift, along with any additional parameters required by the input NL query. A system prompt was used to guide the model's output format, ensuring that each response followed the expected NL–SQL structure. An example entry from the training dataset is shown below:\\\\
\texttt{
  "instruction": "Find galaxies with redshift < 0.137 and g - r color   greater than 0.64.",}\\
\texttt{"output": "SELECT p.objID, p.ra, p.dec, s.z AS redshift, p.u, p.g, p.r, p.i, p.z, p.err\_u, p.err\_g, p.err\_r, p.err\_i, p.err\_z FROM SpecObj AS s JOIN PhotoObj AS p ON s.bestObjID = p.objID WHERE s.class = \textquotesingle GALAXY\textquotesingle~AND s.z < 0.137 AND (p.g - p.r) > 0.64"}\\
\\
We validated the syntactic correctness of the training examples by running all queries through SDSS SkyServer and confirming that they executed without errors, fixing any that failed. We also manually checked most queries for semantic accuracy, ensuring that the SQL queries correctly capture the intent of the natural language input. In total, the dataset contained approximately 2,500 NL--SQL pairs, which was divided into training (90\%) and validation (10\%) subsets during fine-tuning.

Fine-tuning was performed using the HuggingFace \texttt{transformers} library, with LoRA (Low-Rank Adaptation) adapters technique for parameter-efficient training \citep{hu22}. In LoRA, the model is trained on a small subset (low-rank) of the pre-trained weights rather than updating all model parameters, greatly reducing memory and computation requirements. In this work, we used a rank of $r = 8$, a scaling factor of $\alpha = 16$, and a dropout of 0.05 applied to the LoRA updates. Input sequences were truncated or padded to a maximum length of 512 tokens, sufficient to capture most query contexts. Training was conducted with a batch size of 1 and ran for 5 epochs. Default values were used for other training arguments, including the learning rate ($5 \times 10^{-5}$), AdamW optimizer settings, and evaluation strategy (\texttt{eval\_strategy=\textquotesingle epoch\textquotesingle}). Training was conducted on a single NVIDIA T4 GPU with 16 GB memory and it required approximately 40 minutes. After training, the model was quantized to allow deployment on CPUs, making it practical for use on laptops and lightweight environments such as Ollama. This ensures that the system can be used without access to high-performance computing resources.

For evaluation, we considered both syntactic and semantic performance. Syntactic accuracy measures whether the generated SQL queries are valid and executable by running them through SDSS SkyServer and checking for errors, while semantic accuracy was assessed by manual inspection, which evaluates whether the generated SQL query correctly captures the intent of the natural language instruction.

\section{Results}

The fine-tuned Phi-2 model demonstrated strong performance for a lightweight model. Across the validation set, syntactic accuracy reached $\sim 94$\%, indicating that nearly all generated queries were structurally valid SQL. The semantic accuracy was lower, at 60--70\%. For example, when prompted with a NL request ``Find galaxies with redshift between 0.0168 and 0.0229, g - r $>$ 0.48, and r-band magnitude $<$ 19.82.'', the model produced the following SQL:
\begin{lstlisting}
SELECT p.objID, p.ra, p.dec, s.z AS redshift, p.u, p.g, p.r, p.i, p.z FROM PhotoObj AS p JOIN SpecObj AS s ON s.bestObjID = p.objID WHERE s.class = 'GALAXY' AND s.z BETWEEN 0.0168 AND 0.0229 AND (p.g - p.r) > 0.48 AND p.r < 19.82
\end{lstlisting}
which is both a syntactically and semantically correct output demonstrating the model's ability to incorporate domain-specific functions. However, the model occasionally struggled to produce correct results in cases where multiple conditions and ranges of parameter values were provided. For example, the model sometimes hallucinated by generating incorrect parameter ranges or by producing a range when the input specified only a `greater than' or `less than' limit. It reflects the challenges in capturing the subtle details of the interpretation of the schema and complex joins. Some of these problems could arise from unverified examples in the training dataset. Future work could focus on dataset refinement and schema-aware decoding to further reduce such occurrences.

\section{Discussion}

The results highlight the potential of lightweight fine-tuned models in astronomy. By tailoring Microsoft Phi-2 to SDSS, we created a system capable of translating plain-language instructions into valid SQL queries. Importantly, the model achieves this with high syntactic reliability, enabling robust execution, and sufficient semantic accuracy to be useful in many research contexts. Compared to general-purpose LLMs, the fine-tuned Phi-2 offers the advantages of lower resource requirements, offline deployment, and faster inference, making it well-suited for integration into astronomical workflows. We note that experiments with training the model on smaller datasets produced noticeably lower performance, emphasizing the importance of sufficient training examples to capture the diversity of query structures and parameter variations. While the model has a reasonable performance, some errors or inconsistencies may arise. These may reflect unverified NL-SQL pairs in the training set, which could be addressed in future iterations. Using slightly larger models, such as Code LLaMA or equivalent, could further improve performance and reduce such errors.

Beyond technical performance, the most significant impact lies in the scientific use cases. For proposal preparation, researchers often need to select targets based on redshift, brightness, extinction, or morphology. Our system allows such queries to be made in natural language, streamlining the process of building clean samples for JWST, ALMA, or HST. In studies of AGN feedback, complex queries involving morphological parameters (e.g., concentration indices such as R90/R50) and classification flags can be constructed more easily, facilitating large-sample analyses. In educational settings, the system lowers barriers for students who may be new to databases. By enabling them to interact with SDSS through natural language, the tool promotes exploratory learning and engagement without requiring technical training in SQL.

Looking ahead, the approach is scalable to upcoming surveys such as LSST, DESI, SKA and others, where data volumes and complexity will far exceed those of SDSS. Natural language interfaces may become essential tools for exploring petabyte-scale archives, supporting tasks like light curve retrieval, cross-survey source matching, and filtering of alert streams in real time. Integrating ontology-based reasoning, schema explanations, and feedback-driven corrections could further improve usability and robustness.

\section{Conclusion and Future Work}

We have demonstrated that a fine-tuned Microsoft Phi-2 model can serve as an effective natural language interface to SDSS, achieving high syntactic accuracy and promising semantic performance while remaining lightweight and deployable. This work illustrates the feasibility of domain-specific NLIDBs in astronomy and highlights their potential for simplifying access to complex archives.

Future work could explore curating and expanding the training dataset, improving schema awareness, and adding support for additional data types or user feedback mechanisms. While these directions are promising, our current study establishes a proof-of-concept for domain-specific natural language interfaces in astronomy.

\section{Acknowledgment}
This work benefited from SDSS tutorial resources, Lightning.ai training infrastructure, and the HuggingFace open-source ecosystem. We used OpenAI's ChatGPT to assist with code debugging, text paraphrasing, and improving the clarity of the manuscript.

\section{Data Availability}
The fine-tuned Phi-2 model and associated training code are available at:\\
- HuggingFace model: \url{https://huggingface.co/tamhanepd/phi-2-nl-to-sql-sdss}\\
- GitHub repository: \url{https://github.com/tamhanepd/SDSS\_Agent/}

\bibliographystyle{iaulike}
\bibliography{iauguide}

\end{document}